\documentclass[longauth]{aa} 

\usepackage{graphicx}
\usepackage[super]{nth}

\usepackage{txfonts}
\usepackage{siunitx}
\usepackage{tabularx}
\usepackage[switch]{lineno}
\usepackage{subcaption}
\usepackage{lipsum}
\usepackage{multirow}
\usepackage{titlesec}
\usepackage{caption}
\usepackage{amsmath}
\usepackage{titlesec}
\usepackage{caption}
\usepackage{tablefootnote}
\usepackage{xcolor}
\usepackage[utf8]{inputenc}

\usepackage[colorlinks=true, linkcolor=blue, citecolor=blue, urlcolor=blue]{hyperref}
\usepackage{gensymb}

\begin{document}

\title{H.E.S.S. detection and multi-wavelength study of the $z \sim$ 1 blazar PKS 0346$-$27}

\author {
H.E.S.S. Collaboration, F.~Aharonian\inst{\ref{inst1},\ref{inst2}}  
\and M.~Backes\inst{\ref{inst3},\ref{inst4}}  
\and R.~Batzofin\inst{\ref{inst5}}  
\and Y.~Becherini\inst{\ref{inst6},\ref{inst7}}  
\and D.~Berge\inst{\ref{inst8},\ref{inst9}}  
\and K.~Bernl\"ohr\inst{\ref{inst2}}  
\and B.~Bi\protect\footnotemark[1]\inst{\ref{inst10}}  
\and M.~B\"ottcher\protect\footnotemark[1]\inst{\ref{inst4}}  
\and C.~Boisson\inst{\ref{inst11}}  
\and J.~Bolmont\inst{\ref{inst12}}  
\and F.~Brun\inst{\ref{inst13}}  
\and B.~Bruno\inst{\ref{inst14}}  
\and C.~Burger-Scheidlin\inst{\ref{inst1}}  
\and D.~Cecchin~Momesso\inst{\ref{inst14}}  
\and J.~Celic\inst{\ref{inst14}}  
\and M.~Cerruti\inst{\ref{inst6}}  
\and A.~Chen\inst{\ref{inst15}}  
\and M.~Chernyakova\inst{\ref{inst16},\ref{inst1}}  
\and J. O.~Chibueze\inst{\ref{inst4},\ref{inst3}}  
\and O.~Chibueze\protect\footnotemark[1]\inst{\ref{inst4}}  
\and B.~Cornejo\inst{\ref{inst13}}  
\and G.~Cotter\inst{\ref{inst17}}  
\and J.~Damascene~Mbarubucyeye\inst{\ref{inst8}}  
\and I.D.~Davids\inst{\ref{inst3}}  
\and J.~de~Assis~Scarpin\inst{\ref{inst18}}  
\and M.~de~Bony~de~Lavergne\inst{\ref{inst13},\ref{inst19}}  
\and M.~de~Naurois\inst{\ref{inst18}}  
\and E.~de~O\~na~Wilhelmi\inst{\ref{inst8}}  
\and A.~G.~Delgado~Giler\inst{\ref{inst9}}  
\and J.~Devin\inst{\ref{inst20}}  
\and A.~Djannati-Ata\"i\inst{\ref{inst6}}  
\and A.~Dmytriiev\inst{\ref{inst4}}  
\and K.~Egberts\inst{\ref{inst5}}  
\and K.~Egg\inst{\ref{inst14}}  
\and J.-P.~Ernenwein\inst{\ref{inst19}}  
\and C.~Esca~{n}uela~Nieves\inst{\ref{inst2}}  
\and K.~Feijen\inst{\ref{inst6}}  
\and M.~D.~Filipovic\inst{\ref{inst21}}  
\and G.~Fontaine\inst{\ref{inst18}}  
\and S.~Funk\inst{\ref{inst14}}  
\and S.~Gabici\inst{\ref{inst6}}  
\and M.~Genaro\inst{\ref{inst14}}  
\and J.F.~Glicenstein\inst{\ref{inst13}}  
\and J.~Glombitza\inst{\ref{inst14}}  
\and P.~Goswami\inst{\ref{inst22}}  
\and L.~Heckmann\inst{\ref{inst6}}  
\and B.~Hess\inst{\ref{inst10}}  
\and J.A.~Hinton\inst{\ref{inst2}}  
\and W.~Hofmann\inst{\ref{inst2}}  
\and T.~L.~Holch\inst{\ref{inst8}}  
\and M.~Holler\inst{\ref{inst23}}  
\and D.~Horns\inst{\ref{inst24}}  
\and M.~Jamrozy\inst{\ref{inst25}}  
\and F.~Jankowsky\protect\footnotemark[1]\inst{\ref{inst22}}  
\and I.~Jaroschewski\inst{\ref{inst13}}  
\and I.~Jung-Richardt\inst{\ref{inst14}}  
\and K.~Kasprzak\inst{\ref{inst25}}  
\and K.~Katarzy\'{n}ski\inst{\ref{inst26}}  
\and D.~Kerszberg\inst{\ref{inst12}}  
\and B. Khélifi\inst{\ref{inst6}}  
\and N.~Komin\inst{\ref{inst20},\ref{inst15}}  
\and K.~Kosack\inst{\ref{inst13}}  
\and D.~Kostunin\inst{\ref{inst8}}  
\and R.G.~Lang\inst{\ref{inst14}}  
\and S.~Lazarevi\'c\inst{\ref{inst21},\ref{inst27}}  
\and A.~Lemi\`ere\inst{\ref{inst6}}  
\and J.-P.~Lenain\inst{\ref{inst12}}  
\and P.~Liniewicz\inst{\ref{inst25}}  
\and A.~Luashvili\protect\footnotemark[1]\inst{\ref{inst4}}  
\and J.~Mackey\inst{\ref{inst1}}  
\and D.~Malyshev\inst{\ref{inst10}}  
\and V.~Marandon\inst{\ref{inst13}}  
\and M.~Mayer\inst{\ref{inst14}}  
\and A.~Mehta\inst{\ref{inst8}}  
\and A.M.W.~Mitchell\inst{\ref{inst14}}  
\and R.~Moderski\inst{\ref{inst28}}  
\and L.~Mohrmann\inst{\ref{inst2}}  
\and A.~Montanari\inst{\ref{inst22}}  
\and E.~Moulin\inst{\ref{inst13}}  
\and J.~Niemiec\inst{\ref{inst29}}  
\and M.O.~Moghadam\inst{\ref{inst5}}  
\and S.~Panny\inst{\ref{inst23}}  
\and R.D.~Parsons\inst{\ref{inst9}}  
\and U.~Pensec\inst{\ref{inst12}}  
\and P.~Pichard\inst{\ref{inst6}}  
\and T.~Preis\inst{\ref{inst23}}  
\and G.~P\"uhlhofer\inst{\ref{inst10}}  
\and M.~Punch\inst{\ref{inst6}}  
\and A.~Quirrenbach\inst{\ref{inst22}}  
\and A.~Reimer\inst{\ref{inst23}}  
\and O.~Reimer\inst{\ref{inst23}}  
\and I.~Reis\inst{\ref{inst13}}  
\and B.~Rudak\inst{\ref{inst28}}  
\and K.~Sabri\inst{\ref{inst20}}  
\and V.~Sahakian\inst{\ref{inst30}}  
\and D. Jimeno\inst{\ref{inst8}}  
\and A.~Santangelo\inst{\ref{inst10}}  
\and M.~Sasaki\inst{\ref{inst14}}  
\and F.~Sch\"ussler\inst{\ref{inst13}}  
\and J.N.S.~Shapopi\inst{\ref{inst3}}  
\and W.~Si~Said\inst{\ref{inst18}}  
\and {\L.}~Stawarz\inst{\ref{inst25}}  
\and S.~Steinmassl\inst{\ref{inst2}}  
\and T.~Takahashi\inst{\ref{inst31}}  
\and T.~Tanaka\inst{\ref{inst32}}  
\and A.M.~Taylor\inst{\ref{inst8}}  
\and G.~L.~Taylor\inst{\ref{inst22}}  
\and R.~Terrier\inst{\ref{inst6}}  
\and T.~Unbehaun\inst{\ref{inst14}}  
\and C.~van~Eldik\inst{\ref{inst14}}  
\and M.~Vecchi\inst{\ref{inst33}}  
\and C.~Venter\inst{\ref{inst4}}  
\and J.~Vink\inst{\ref{inst34}}  
\and T.~Wach\inst{\ref{inst14}}  
\and S.J.~Wagner\inst{\ref{inst22}}  
\and A.~Wierzcholska\inst{\ref{inst29},\ref{inst22}}  
\and M.~Zacharias\inst{\ref{inst22},\ref{inst4}}  
\and A.~Zech\inst{\ref{inst11}}  
\and W.~Zhong\inst{\ref{inst8}}  
}

\institute{
Astronomy \& Astrophysics Section, School of Cosmic Physics, Dublin Institute for Advanced Studies, DIAS Dunsink Observatory, Dublin D15 XR2R, Ireland\label{inst1}
\and Max-Planck-Institut für Kernphysik, P.O. Box 103980, D 69029 Heidelberg, Germany\label{inst2}
\and University of Namibia, Department of Physics, Private Bag 13301, Windhoek 10005, Namibia\label{inst3}
\and Centre for Space Research, North-West University, Potchefstroom 2520, South Africa\label{inst4}
\and Institut für Physik und Astronomie, Universit\"at Potsdam, Karl-Liebknecht-Strasse 24/25, D 14476 Potsdam, Germany\label{inst5}
\and Universit\'e Paris Cit\'e, CNRS, Astroparticule et Cosmologie, F-75013 Paris, France\label{inst6}
\and Department of Physics and Electrical Engineering, Linnaeus University, 351 95 V\"axj\"o, Sweden\label{inst7}
\and Deutsches Elektronen-Synchrotron DESY, Platanenallee 6, 15738 Zeuthen, Germany\label{inst8}
\and Institut f\"ur Physik, Humboldt-Universit\"at zu Berlin, Newtonstr. 15, D 12489 Berlin, Germany\label{inst9}
\and Institut f\"ur Astronomie und Astrophysik, Universit\"at T\"ubingen, Sand 1, D 72076 T\"ubingen, Germany\label{inst10}
\and LUX, Observatoire de Paris, Universit\'e PSL, CNRS, Sorbonne Universit\'e, 5 Pl. Jules Janssen, 92190 Meudon, France\label{inst11}
\and Sorbonne Universit\'e, CNRS/IN2P3, Laboratoire de Physique Nucl\'eaire, et de Hautes Energies, LPNHE, 4 place Jussieu, 75005 Paris, France\label{inst12}
\and IRFU, CEA, Universit\'e Paris-Saclay, F-91191 Gif-sur-Yvette, France\label{inst13}
\and Friedrich-Alexander-Universit\"at Erlangen-N\"urnberg, Erlangen Centre for Astroparticle Physics,  Nikolaus-Fiebiger-Str. 2, 91058 Erlangen, Germany\label{inst14}
\and School of Physics, University of the Witwatersrand, 1 Jan Smuts Avenue, Braamfontein, Johannesburg, 2050, South Africa\label{inst15}
\and School of Physical Sciences and Centre for Astrophysics \& Relativity, Dublin City University, Glasnevin, Dublin D09 W6Y4, Ireland\label{inst16}
\and University of Oxford, Department of Physics, Denys Wilkinson Building, Keble Road, Oxford OX1 3RH,  UK\label{inst17}
\and Laboratoire Leprince-Ringuet, \'Ecole Polytechnique, CNRS, Institut Polytechnique de Paris, F-91128 Palaiseau, France\label{inst18}
\and Aix Marseille Universit\'e, CNRS/IN2P3, CPPM, Marseille, France\label{inst19}
\and Laboratoire Univers et Particules de Montpellier, Universit\'e Montpellier, CNRS/IN2P3, CC 72, Place Eug\`ene Bataillon, F-34095 Montpellier Cedex 5, France\label{inst20}
\and School of Science, Western Sydney University, Locked Bag 1797, Penrith South DC, NSW 2751, Australia\label{inst21}
\and Landessternwarte, Universit\"at Heidelberg, K\"onigstuhl, D 69117 Heidelberg, Germany\label{inst22}
\and Universit\"at Innsbruck, Institut f\"ur Astro- und Teilchenphysik, Technikerstraße 25, 6020 Innsbruck, Austria\label{inst23}
\and Universit\"at Hamburg, Institut f\"ur Experimentalphysik, Luruper Chaussee 149, D 22761 Hamburg, Germany\label{inst24}
\and Obserwatorium Astronomiczne, Uniwersytet Jagiello\'nski, ul. Orla 171, 30-244 Krak\'ow, Poland\label{inst25}
\and Institute of Astronomy, Faculty of Physics, Astronomy and Informatics, Nicolaus Copernicus University, Grudziadzka 5, 87-100 Torun, Poland\label{inst26}
\and Associate Member\label{inst27}
\and Nicolaus Copernicus Astronomical Center, Polish Academy of Sciences, ul. Bartycka 18, 00-716 Warsaw, Poland\label{inst28}
\and Instytut Fizyki Jadrowej PAN, ul. Radzikowskiego 152, ul. Radzikowskiego 152, 31-342 Krak\'ow, Poland\label{inst29}
\and Yerevan Physics Institute, 2 Alikhanian Brothers St., 0036 Yerevan, Armenia\label{inst30}
\and Kavli Institute for the Physics and Mathematics of the Universe (WPI), The University of Tokyo Institutes for Advanced Study (UTIAS), Japan\label{inst31}
\and Department of Physics, Konan University, 8-9-1 Okamoto, Higashinada, Kobe, Hyogo 658-8501, Japan\label{inst32}
\and Kapteyn Astronomical Institute, University of Groningen, Landleven 12, 9747 AD Groningen, The Netherlands\label{inst33}
\and GRAPPA, Anton Pannekoek Institute for Astronomy, University of Amsterdam, Science Park 904, 1098 XH Amsterdam, The Netherlands\label{inst34}
}
\offprints{H.E.S.S.~collaboration,
    \protect\\\email{\href{mailto:contact.hess@hess-experiment.eu}{contact.hess@hess-experiment.eu}};
    \protect\\\protect\footnotemark[1] Corresponding authors
    }

\abstract
{PKS 0346-27 is a Low Synchrotron Peaked (LSP) blazar at redshift 0.991.  The very-high-energy (VHE, E > 100 GeV) spectra of blazars are always affected by $\gamma\gamma$ absorption by the Extragalactic Background Light (EBL) and subsequently, no blazars have been detected in VHE $\gamma$-rays at redshifts exceeding 1.}
  {Extending the redshift range of VHE-detected blazars to $z \gtrsim 1$ will yield insights into the cosmological evolution of both the VHE blazar population and the EBL. This is the goal of a target-of-opportunity (ToO) programme by H.E.S.S.  to observe flaring high-redshift ($z \gtrsim 1$) blazars.}
 {We report on H.E.S.S. ToO and multi-wavelength observations of the blazar PKS\,0346$-$27. Along with H.E.S.S., simultaneous data from {\it Fermi}-LAT, {\it Swift} (XRT and UVOT), and ATOM have been analysed and modelled using single-zone leptonic and hadronic models.}
 {PKS~0346-27 has been detected by H.E.S.S {Extending the redshift range of VHE-detected blazars to $z \gtrsim 1$ will yield insights into the cosmological evolution of both the VHE blazar population and the EBL. This is the goal of a target-of-opportunity (ToO) programme by H.E.S.S.  to observe flaring high-redshift ($z \gtrsim 1$) blazars.}. at a significance of 6.3$\sigma$ during one night, on 3 November 2021, while for other nights before and after this day, upper limits on the
VHE flux are determined. No evidence for intra-night $\gamma$-ray variability has been found. A flare in high-energy (HE, $E > 100$~MeV) $\gamma$-rays detected by {\it Fermi}-LAT preceded the H.E.S.S. detection by 2 days. A fit with a single-zone emission model to the contemporaneous spectral energy distribution during the detection night was possible with a proton-synchrotron-dominated hadronic model, requiring a proton-kinetic-energy-dominated jet power temporarily exceeding the source's Eddington limit, although alternative (e.g. multi-zone) models can not be ruled out.  A one-zone leptonic model is, in principle, also able  to fit the flare-state SED, however, requiring implausible parameter choices, in  particular, extreme Doppler and bulk Lorentz factors of $\gtrsim 80$.} {}


\maketitle
%

\section{Introduction}   
Blazars are the relativistically boosted class of radio-loud Active Galactic Nuclei (AGN) with their jets aligned at a small angle to the observer’s line of sight \citep{urry1995unified}. Based on their optical spectra, they are classified into two sub-classes, namely: flat-spectrum radio quasars (FRSQs) which have broad emission lines and BL~Lacertae objects (BL Lacs) with weak or no emission lines \citep{stocke1991einstein,sambruna1996spectral}. The broadband spectral energy distributions (SEDs) of blazars show a typical double-peaked structure peaking in the infrared (IR) to X-rays for the low-energy component and in the MeV to TeV bands for the high energy component. The low-energy component is well explained by synchrotron emission from relativistic electrons in the jet while the emission processes giving rise to the high-energy emission peak are not fully understood yet, as several emission mechanisms could be responsible. In the leptonic model, the high energy gamma-ray radiation results from the Inverse-Compton (IC) scattering of soft target photons originating in the synchrotron radiation process  \citep{sikora2009constraining} or external photon fields \citep[e.g.][]{dermer1992high}. Alternatively, in hadronic models  \citep{mucke2001proton,bottcher2013leptonic}, part of the high energy emission originates from proton synchrotron radiation or the decay products following photo-pion interactions of relativistic protons. 
\\
\indent Based on the location of their synchrotron peak frequency ($\nu_{peak,sync}$), blazars are also classified into three categories, known as low-synchrotron-peaked (LSP) blazars, which have $\nu_{peak,sync} < 10^{14}$~Hz, intermediate-synchrotron-peaked (ISP) blazars, which have $10^{14}$~Hz $\le \nu_{peak,sync} \le 10^{15}$~Hz  and high-synchrotron-peaked (HSP) blazars with $\nu_{peak,sync} > 10^{15}$~Hz \citep{Abdo10blazarSEDs}. Occasionally, LSP sources transition to an ISP character during their flaring state \citep{foschini2008infrared,cutini2014radio,Ahnen15_PKS1441}. \\
\indent The $\gamma\gamma$ pair production through the interaction of $\gamma$-ray photons with low-energy photons (infrared and optical) from diffuse extragalactic background light (EBL) limits the possibility of detecting blazars above 100 GeV from sources at cosmological distances \citep{salamon1998absorption}.
This absorption effect increases with distance and energy of the VHE photons. Therefore VHE blazars (usually not detected beyond redshift $\sim 0.5$) at redshift $\sim 1$ are expected to be severely affected by EBL absorption. However, this absorption can be used to constrain this EBL effect especially for high redshift sources. With this science objective, the High Energy Stereoscopic System (H.E.S.S.) collaboration has a target-of-opportunity (ToO) program in place to trigger H.E.S.S. and coordinated multi-wavelength observations of high redshift ($z \gtrsim 1$) blazars in flaring states identified by the {\it Fermi} Large Area Telescope \citep{2024icrc.confE.924C}. 
 \\
\indent The FSRQ PKS 0346-27 
 was first identified as a radio source in the Parkes catalog \citep{bolton1964parkes} and later classified as a quasar by \cite{white1988redshifts} based on its optical spectrum. It was detected as an X-ray source by ROSAT \citep{voges1999rosat} and as a $\gamma$-ray source by {\it Fermi}-LAT and included in the {\it Fermi}-LAT First Source catalog \citep[1FGL;][]{abdo2010fermi}. In the {\it Fermi} -LAT fourth source catalog \citep[4FGL;][]{2020ApJS..247...33A} it is associated with the $\gamma$-ray source 4FGL J0348.5-2749. 
On 2 February 2018 (MJD 58151), \cite{angioni2018fermi} reported a strong $\gamma$-ray flare based on {\it Fermi}-LAT data. 
This prompted multi-wavelength follow-up observations in the optical -- Near-Infrared (NIR)  \citep[] {nesci2018high, vallely2018asas}, ultraviolet (UV) and X-rays \citep{nesci2018x} which also showed enhanced activity of this source. \cite{kamaram2023multifrequency} studied the long-term variability of PKS 0346-27 from 2019 December to 2021 January (MJD 58484-59575) with multi-wavelength data from {\it Fermi}-LAT, {\it Swift}-XRT and UVOT and found five flaring episodes during this period of study with a minimum variability time scale of $\sim 1$~day and $\sim 0.1$ days for the $\gamma$-ray and X-ray light curves, respectively.
\\
\indent In this paper, we 
report on the first VHE $\gamma$-ray detection of this source by H.E.S.S. on 3 November 2021 (MJD 59521.97) from observations triggered by a $\gamma$-ray flare detected by {\it Fermi}-LAT. Contemporaneous multi-wavelength data from H.E.S.S, {\it Fermi}-LAT, ATOM, {\it Swift}-XRT and UVOT during this flare are presented. 
We modelled the SED of this source using the leptonic and hadronic code by \cite{bottcher2013leptonic} and constrain the effects of  EBL using the EBL model by \cite{finke2010modeling}. In Section 2, we discuss the multi-wavelength observations and analysis techniques for the different telescopes and present the multi-wavelength SED and lightcurves. The SED modelling of the source is discussed in Section 3. We present a summary and final conclusions in Section 4.

\section{Observation and data analysis}
\subsection{H.E.S.S. observations and analysis techniques}
\begin{figure*}
    \centering
    \includegraphics[width=0.8\textwidth]{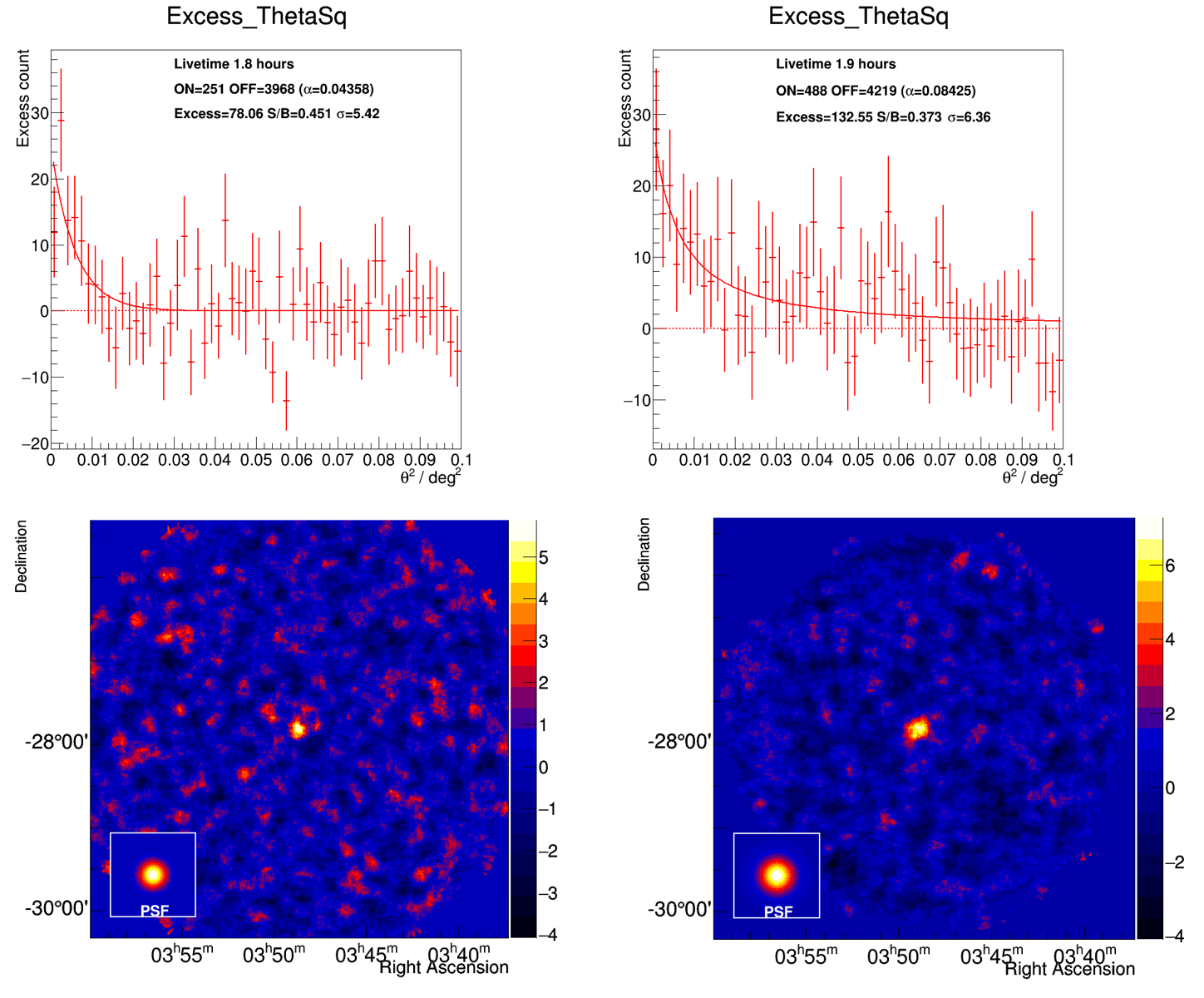}
    \caption{November 3, 2021 (MJD 59521.93 - 59521.99) observation night analysis result for $\theta^2$ and significance plots with the 4 selected runs. Top and bottom left are results from CT1-4 data set while top and bottom right are results from CT5 data set. The inset shows the Point spread function (PSF) which is derived by fitting the $\theta^2$ distribution to the KING's function, described in \cite{2013ApJ...765...54A}}
    \label{fig:combined}
\end{figure*}

The H.E.S.S. array consists of five telescopes located in the Southern Hemisphere in Namibia at an altitude of 1800~m. It detects VHE $\gamma$-rays using the atmospheric Cherenkov imaging technique \citep{2015ICRC...34..847V}. In 2003, the first phase consisting of four 12m diameter telescopes (CT1-4), arranged on a 120m side square, started operations. It was sensitive to $\gamma$-rays above a few hundred GeV.  In 2012, an additional 28m telescope (CT5) was added in the middle of the four 12m telescopes, which lowered the detection threshold of the array.\\
\indent H.E.S.S. observation data is collected in observation runs of approximately 28 mins. The data taken with the four 12m telescopes (CT1-4) is referred to as `stereo' data, while the data collected with only CT5 is called `mono' data. Each telescope images the Cherenkov light emitted by particle showers triggered by the
interaction of $\gamma$-rays or cosmic rays with the Earth’s atmosphere. A probabilistic distinction  between cosmic-ray and $\gamma$-ray induced showers is done based on the characteristics of the shower images. The study of the shower images allows reconstruction of the primary particle's direction, energy, and arrival time. 
\\
\indent The ToO observations of PKS 0346-27 with H.E.S.S. were triggered on 30 October 2021 (MJD 59517.97) by FlaapLUC \citep{Lenain_2018} following the detection of flaring activity by {\it Fermi}-LAT. H.E.S.S. observations were carried out with all five telescopes during the observation period (30 October - 8 November 2021, MJD 59517.97 - 59525.94). 
The observations were done in wobble mode, where the telescopes point at 0.5$^{\circ}$  from the source position to allow simultaneous background estimation \citep{2007A&A...466.1219B}. Data quality cuts were applied to remove periods affected by poor weather conditions and hardware problems. After applying the quality criteria, 34 runs were selected for the entire observation period, out of which 4 runs were selected for the detection night (MJD 59521.97). 
\\
\indent The H.E.S.S. data for all the observation periods have been analysed with the IMPACT (Image Pixel-wise fit for Atmospheric Cherenkov Telescope)  analysis chain \citep{2014APh....56...26P} for only CT1-4 data. The analysis was cross-checked with another analysis chain \citep{2009APh....32..231D} and this yielded a consistent result. Another independent analysis was performed with only the CT5 data set using the neural-network-based chain described in \citet{2015ICRC...34.1022M}.
\\
\indent The CT1-4 analysis of the full data set yielded an excess of 50.8 $\gamma$-like events and a signal-to-background ratio of 0.03 for 16.6 h livetime. Using Equ. (17) of \cite{1983ApJ...272..317L} a significance of 1.3$\sigma$  was obtained for the entire dataset after background subtraction using the ring background method \citep{2006A&A...457..899A} which resulted in 1618 ON and 35372 OFF events, while the averaged number of OFF regions $\alpha$ is 9.24. The signal-to-background ratio is defined as excess / background where background (B) = (\# OFF events)/$\alpha$. The CT5 analysis yielded an excess of 208.1 $\gamma$-like events corresponding to 3.6$\sigma$ for 16.8 h livetime. Analysing individual nights, only one night (3 November, 2021) showed a significant greater than 5$\sigma$ detection. Therefore, upper limits were generated for both spectral and light curve analyses for all other observation nights at 99~\%  confidence level following \citet{rolke2005limits}. 
\\
 \indent Figure \ref{fig:combined} shows the sky maps and the on-source and normalized off-regions distributions as a function of squared angular distance from the source (i.e. $\theta^2$  plots) for the detection night for both analysis chains. The 4 runs passed the quality selection criteria. The 
  analysis yielded an excess of 78.0 $\gamma$-like events and a signal-to-background ratio of 0.4 for 1.8 h  livetime, resulting in a significance of 5.4$\sigma$ for CT1-4 while for CT5, it generated an excess of 132.5 $\gamma$-like events for 1.9 h livetime and a significance of 6.3$\sigma$. The ring background method was used for background subtraction, which resulted in 251 ON, 3968 OFF events for CT1-4 and 488 ON, 4219 OFF events for CT5.                
The energy threshold for CT1-4 is 133 GeV and 121 GeV for CT5.  
Only the detection night analysis was used for the spectral modelling while all observation nights were used to generate the H.E.S.S. light curve, using the photon index from  detection night. The reflected background method \citep{2006A&A...457..899A} was used for the spectral analysis for both analysis chains. The spectrum was fitted with a power-law function:
 \begin{equation}
     \frac{\mathrm{d}N}{\mathrm{d}E} = N_0 \left(\frac{E}{E_0}\right)^{-\Gamma},
 \end{equation}
  where $N_0$ is the differential photon flux normalization at the reference energy $E_0$, and  $\Gamma$ is the photon index.

The systematic errors were derived separately, following \cite{hess2022Sci}. Multiple sources of systematic errors were taken into account. For the CT1-4 analysis, the systematic uncertainties derived from the different analysis cuts are 5\% on the flux normalization and 0.1 on the index. The atmospheric  transparency contributes
a relative uncertainty of 10\% on the flux normalization and 0.05 on the index. For the CT5 analysis, the systematic error due to the imperfect background acceptance description is 15\% on the flux normalization and 0.15 on the index, while the atmospheric transparency contributes 7\% on the flux normalization and 0.15 on the index, respectively. The systematic uncertainty on the flux normalization derived from the uncertainty of the absolute energy scale, which is assumed to be 10\%, is estimated as $(1 \pm 10 \%)^{\Gamma}$. All these contributions are added in quadrature in order to obtain the overall systematic errors which are listed in Tab. \ref{tab:fit_error} with spectral parameters.

\begin{figure}[htp]
  \begin{subfigure}[b]{0.5\textwidth}
    \includegraphics[width=\textwidth]{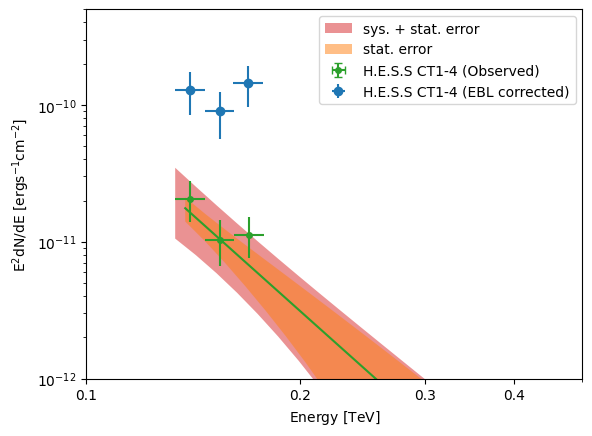}
    \caption{Observed spectrum and flux points together with de-absorbed flux points for CT1-4 data set}
    \label{fig:f1}
  \end{subfigure}
  \begin{subfigure}[b]{0.5\textwidth}

  \includegraphics[width=\textwidth]{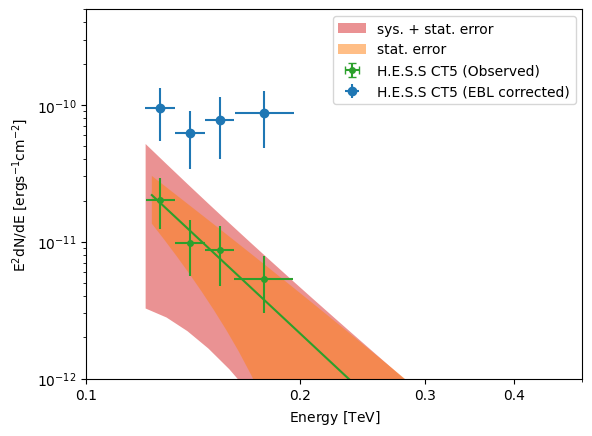}
    \caption{Observed spectrum and flux points together with de-absorbed flux points  for CT5 data set}
    \label{fig:f2}
  \end{subfigure}
  \caption{Observed spectra and flux points together with intrinsic flux points for CT1-4 (Figure a) and CT5 (Figure b). The observed spectrum and flux points were de-absorbed with \cite{finke2010modeling} EBL model. The orange bands are the statistical errors only while the purple bands show the systematic together with statistical errors. The error bars of the flux points are statistical only.}
  \label{fig:f3}
\end{figure}

\indent Figure \ref{fig:f3} shows the energy spectrum where each flux point is rebinned to achieve a minimum significance of 2$\sigma$ per bin (forward-folding method). The butterfly-shaped areas indicate the statistical errors (orange) and systematic together with statistical errors (purple). Fig. \ref{fig:f3} shows both the observed and EBL corrected flux points using the model of \cite{finke2010modeling}.

\begin{table*}
\caption{VHE $\gamma$-ray spectral parameters for the detection night of 3 November 2021.}
\label{tab:fit_error}
\begin{center}
\begin{tabular}{cccc} 
 \hline
 \hline
Analysis chain & $E_0$ & $N_0$ & $\Gamma$  \\
 &  [TeV] & [$10^{-10}$\,$cm$\,$^{-2}$s$^{-1}$TeV] & \\
\hline
\vspace{2mm}
CT1-4    &  0.150  & 3.3 $\pm$ 0.79(stat.) $^{+2.9}_{-1.7}$ (syst.) & 6.6 $\pm$ 1.2 (stat.) $\pm$ 0.1(syst.) \\

CT5 &  0.150\tablefootmark{a}  & 2.4 $\pm$ 1.5(stat.) $^{+2.2}_{-1.3}$ (syst.) & 6.8 $\pm$ 1.3 (stat.) $\pm$ 0.2 (syst.) \\
\hline
\end{tabular}
\end{center}
\tablefoottext{a}{The decorrelation energy of CT5 analysis was found at its energy threshold. However, the correlation coefficient at this energy remains significantly large, indicating that the fit parameters are still strongly coupled and the decorrelation is not optimal. The same reference energy as CT1-4 analysis was chosen for CT5 analysis, to compare $N_0$ at the reference energy. The given reference energy is not the decorrelation energy implying larger statistical errors in the spectral parameters.}
\end{table*}

\subsection{{\it Fermi}-LAT data analysis}
The Large Area Telescope (LAT) is the primary instrument onboard the {\it Fermi} satellite \citep{atwood2009large}, sensitive to $\gamma$-rays with energies from $\sim$ 20 MeV to beyond 300 GeV. It enables high sensitivity measurements and operates in full-sky survey mode, covering the entire sky approximately every 3 hours. 
\\
\indent {\it Fermi}-LAT data taken between 28 October and 8 November 2021 (MJD 59515.5 - 59526.5) were retrieved from the {\it Fermi} Science Support Center data server\footnote{\url{https://Fermi.gsfc.nasa.gov/cgi-bin/ssc/LAT/LATDataQuery.cgi}} and analysed using {\it Fermitools} version 2.2.0, along with P8R3$\_$SOURCE$\_$V3 instrument response functions. Only source class events (event class 128 and event type 3) were selected, in the 15\degree\ region of interest (ROI), centered around the position of the point source 4FGL\,J0348.6$-$2749 associated with PKS\,0346$-$27, with energies between 100 MeV and 300 GeV. A maximum zenith angle cut of 90$\degree$  was applied to filter out the Earth's limb contamination. The  \texttt{gll$\_$iem$\_$v07} and \texttt{iso$\_$P8R3$\_$SOURCE$\_$V3$\_$v1} models were used to account for the diffuse Galactic and extra-galactic isotropic background emissions, respectively. 
\\
\indent A binned likelihood analysis was performed in an iterative manner. The initial input source model is composed of all 12-year {\it Fermi}-LAT catalog (4FGL-DR3) sources (\cite{2020ApJS..247...33A}) that lie in the 15\degree\ ROI centered around the position of 4FGL\,J0348.6$-$2749, to which additional 10\degree\ are added by the \textit{make4FGLxml.py} tool to avoid event leakage. Then, parameters of all sources contributing to less than 5\% of the total measured counts and with Test Statistic (TS) below 9, are kept constant. Finally, only parameters of sources located within 3\degree\ separation from the target of interest and which were not frozen in the previous steps are allowed to vary, along with the normalization factors of the two diffuse background models. 
\\
\indent
{\it Fermi}-LAT light curves for the 28 October - 8 November 2021 period, centred around the VHE detection night, with daily and 12 h time bins, were derived. They exhibit strong flux variability, with fractional variability amplitude \citep{Vaughan_2003} $F_{var}$ = (82 $\pm$ 5)~\%. 
A maximum daily-binned flux value ($>$100 MeV) of (3.14 $\pm$ 0.25)\,10$^{-6}\ \rm{ph\ cm^{-2}\ s^{-1}}$ was measured. A comparable (historical) maximum was only announced in April 2019 \citep[ATel\#12693,][]{ATel12693_Apr2019}, when the authors also reported a significant spectral hardening with respect to the first release of the 4FGL (8 year {\it Fermi}-LAT catalog), where the average PL index of the source was 2.43~$\pm$~0.05. Indeed, PKS\,0346$-$27 remained in a persistently very low state until end of 2017 - early 2018, when it started flaring for more than 3 years (according to the {\it Fermi}-LAT Light Curve Repository\footnote{\url{https://Fermi.gsfc.nasa.gov/ssc/data/access/lat/LightCurveRepository/}}). This changing behaviour is reflected in the spectral index and variability amplitude values that were published later, with more recent releases of the 4FGL catalog (e.g. 12-year LAT catalog 4FGL-DR3, covering its brightest $\gamma$-ray states) where the spectral index hardened to 2.08 $\pm$ 0.01 and the variability index of the source increased by a factor of 180. \\
\indent The spectral indices obtained during the considered campaign showed no significant variability, with a mean value of 1.98 $\pm$ 0.22 which appears harder than the aforementioned older catalog values, but is compatible within errors with the more recent 12 year catalog value of 2.08 $\pm$ 0.01.

The highest energy photon ($E \gtrsim 14$~GeV, with at least 95\% chance of being associated with the source of interest) was detected at 03:47:30 UTC on 02 November 2021 ($\sim$ MJD 59520.16), at the peak of the {\it Fermi}-LAT light curve, that is approximately two days before the H.E.S.S. detection. No VHE observations were available on the day of the {\it Fermi}-LAT light-curve peak.

The one-day long time window centered around the H.E.S.S. detection night was analysed separately and yielded a 7.2$\sigma$ detection (TS = 52), when using a PL model. Although the source is characterised by a significant spectral curvature in the 4FGL-DR3 catalog ($\Delta$TS = 347), no curvature was found for the one day-long period centered around the H.E.S.S. detection night. PL index of 1.99 $\pm$ 0.21 was derived, compatible with the 4FGL-DR3 catalog value of 2.08 $\pm$ 0.01, while using LogPar spectral model yields TS = 46, log-parabolic index $\alpha$ = 1.84 $\pm$ 0.32 and curvature parameter ($\beta$ = 0.19 $\pm$ 0.22) compatible with zero.


\subsection{{\it Swift} data analysis}

Simultaneous to H.E.S.S. and {\it Fermi}-LAT, PKS 0346-27 was also observed by the Neil Gehrels observatory, with both the {\it Swift}-X-Ray Telescope (XRT) and the {\it Swift}-UV--optical Telescope (UVOT). The XRT is sensitive within the energy range of 0.2 -- 10.0~keV, while the UVOT is observing at UV -- optical wavelengths of 170 -- 650~nm with V, B, and U filters in the optical and W1, M2, and W2 in the UV. During the observation period (MJD 59517 to 59526), PKS 0346-27 was observed by {\it Swift} on three nights, namely MJD 59521.79 (the H.E.S.S. detection night), MJD 59523.78, and 59525.91, with observation IDs 00038373036, 00038373037, 00038373040, respectively, for a total exposure period of 5.9 ks. Only the detection night was considered for the SED modelling, while all three observation periods were used to construct light curves. 

\subsubsection {\textit{Swift}-XRT}

The XRT data was analysed with available tools on Heasoft software. To reduce the data, we used the standard {\tt XRT-PIPELINE} to produce clean event files, using the calibration version CALDB 20220331. The cleaned event files corresponding to the photon counting (PC) mode were used to produce the source and background spectra, using the {\tt XSELECT} tool. For the source and background extractions, we consider circular regions of 20 arcseconds and 40 arcseconds, respectively.  
\\ 
\indent The ancillary response file (ARF) and redistribution matrix file (RMF) were generated with {\tt xrtmkarf} and {\tt quizcif}, respectively. The source spectrum, background spectrum, RMF and ARF files were then merged using the {\tt grppha} tool, with each bin accommodating 10 counts for {\tt Xspec} analysis.  The spectrum for each observation was fit with an absorbed power-law model and column density, $N{_H}=8.16 \times10^{19}$~cm$^{-2}$ \citep{bekhti2016hi4pi}, resulting in photon indices of 1.74 $\pm$ 0.23, 1.87 $\pm$ 0.29 and 2.2 $\pm$ 0.28, respectively.

\subsubsection{\textit{Swift}-UVOT}

PKS 0346-27 was also observed by {\it Swift}-UVOT in the same observation periods as {\it Swift}-XRT for a total observation duration of 5.8~ks. For the spectral and lightcurve analysis, the {\tt uvotsource} tool has been used to extract the magnitudes from the images taken with each filter (V, B, U, UVW1, UVM2, UVW2) considering source and background regions with radii of 5 arcsecs and 10 arcsecs, respectively. The observed magnitudes were corrected for Galactic extinction with $E(B-V) = 0.0094$~mag \citep{schlafly2011measuring} and the extinction ratios for each filter from \cite{giommi2006swift}. The magnitudes were converted to flux points with the photometric zero points from \cite{2011AIPC.1358..373B} and the conversion factors provided by \cite{giommi2006swift}. 
\\

\subsubsection{ATOM}\setlength{\parskip}{0pt}
The Automatic Telescope for Optical Monitoring (ATOM) is a 75~cm telescope located at the H.E.S.S. site in Namibia \citep{atom_hauser}.
ATOM monitored PKS 0346-27 on a daily cadence in Bessel B and R filters, with denser coverage and additional V band observations during several nights.
The ATOM data were analysed using the fully automated ATOM Data Reduction and Analysis Software and their quality has been checked manually.
Absolute flux was obtained via differential photometry using up to six custom-calibrated comparison stars in the same field-of-view.
The results were extinction-corrected following \citet{schlafly2011measuring}.

\section{Results and discussion}

\subsection{Multi-wavelength light curves}

\begin{figure}[h]
    \centering
    \includegraphics[width=\columnwidth]{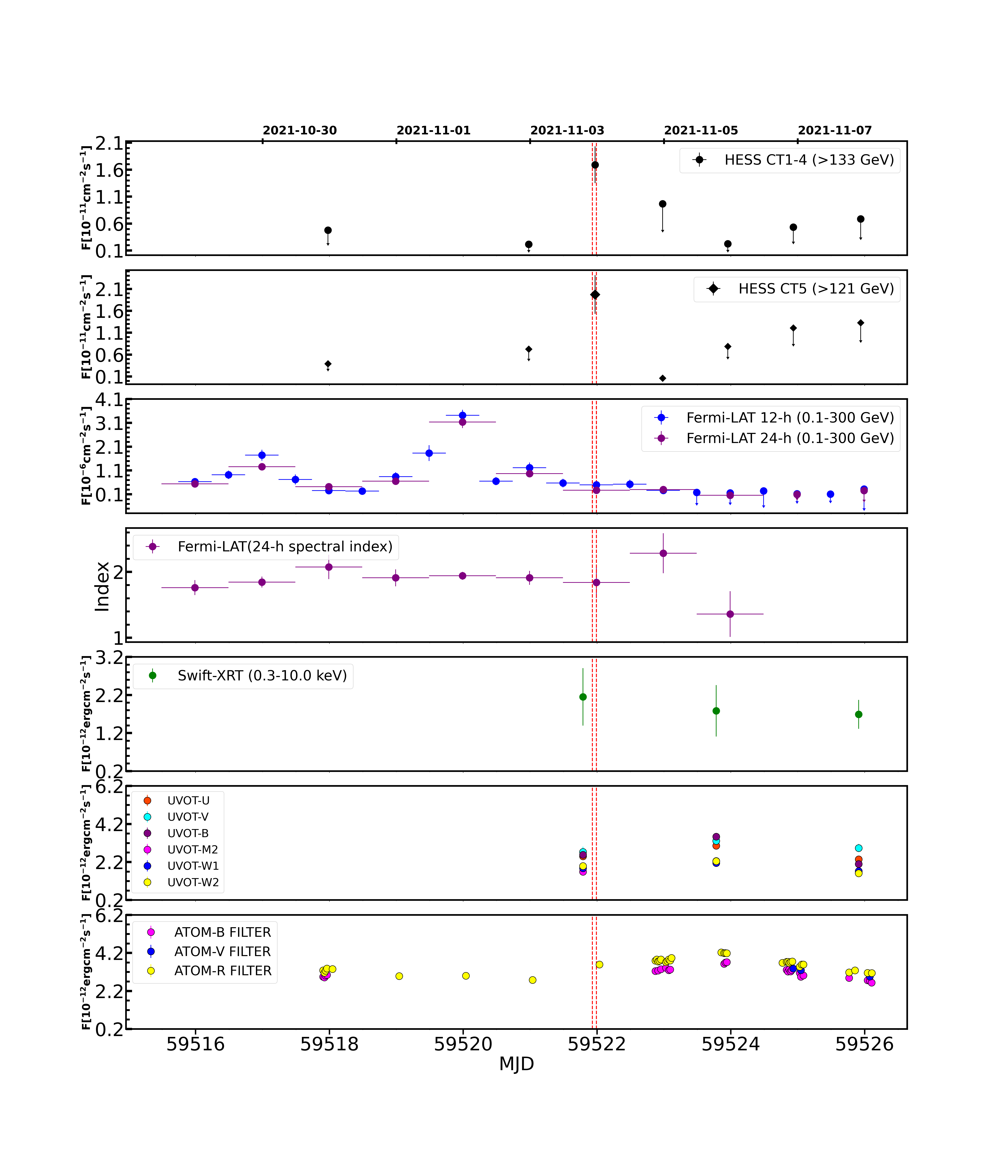}
    \captionsetup{aboveskip=-7pt}
    \caption{Multi-wavelength light curves from MJD 59515.99 - 59526.10. Panels from top to bottom:  H.E.S.S flux for CT1-4 and CT5 in $10$$^{-11}$~cm$^{-2}$~s$^{-1}$, {\it Fermi}-LAT flux for both 12-h and 24-h binning in $10^{-6}$~cm$^{-2}$~s$^{-1}$,  {\it Fermi}-LAT spectral index for 24-h binning, {\it Swift}-XRT flux in $10^{-12}$~erg~cm$^{-2}$~s$^{-1}$, {\it Swift}-UVOT flux in  $10^{-12}$~erg~cm$^{-2}$~s$^{-1}$ and ATOM flux for V, B and R filters in $10^{-12}$~erg~cm$^{-2}$~s$^{-1}$. The double vertical red dotted lines denote the H.E.S.S. detection period, MJD 59521.93 - 59521.99. Only statistical errors are shown.}
    \label{fig:mwl_lc}
\end{figure}

The multi-wavelength light curve 
for the entire observation period from 30th of October (MJD 59517.97) to 8th of November 2021 (MJD 59525.94) for all the telescopes is shown in Fig \ref{fig:mwl_lc}. On the H.E.S.S. detection night (MJD 59521.97), {\it Swift} observed the source 5 hours before, while ATOM observed an hour after H.E.S.S. The VHE $\gamma$-ray night-wise light curves using CT1-4 and CT5 data are shown in the first and second panels                  
while the HE $\gamma$-ray light curves displayed  
in the third panel were obtained for daily and 12 hr binnings, respectively. As seen in the {\it Fermi}-LAT light curve, the source exhibits a strong flux variability and a peak flux of $(3.1\pm0.25) \times 10^{-6}$~ph~cm$^{-2}$~s$^{-1}$ on 1st November, 2021 (MJD 59519.9). 
There was no significant flaring activity seen in both {\it Swift} XRT and UVOT during the HE.S.S flaring period, although there was a flux increase in the UVOT bands: For example, in the U-band, from the first observation night (MJD 59521.79) with a flux of $(2.3 \pm 0.10) \times 10^{-12} $~erg~cm$^{-2}$~s$^{-1}$, it increased to  $(3.0 \pm 0.14) \times 10^{-12} $~erg~cm$^{-2}$~s$^{-1}$ on 5th of November (MJD 59523.78), i.e. 2 days after the H.E.S.S. flare detection. This increased flux is also observed in other filters. There is also an obvious flux increase in the ATOM R filter on 4th November, 2021 (MJD 59522.03) with flux of  $(3.6 \pm 0.13) \times 10^{-12} $~erg~cm$^{-2}$~s$^{-1}$, almost during the time of the H.E.S.S. flare. This flaring activity in ATOM continued until 5th November, 2021 (MJD 59523.85) with a peak flux of $(4.2 \pm 0.10) \times 10^{-12} $~erg~cm$^{-2}$~s$^{-1}$ and gradually reduced until the last observation night.
\\ 
\indent Most notably, the VHE flare appears to be delayed by about 2 days with respect to the GeV flare; the GeV flare had already subsided during the H.E.S.S. flaring period. However, there was no observation in the VHE band during the HE flaring period, so a simultaneous HE - VHE flare during that night cannot be excluded. Delayed flares between the HE and VHE energy bands have previously been observed in 3C 279 \citep{2019ICRC...36..668E} and also between VHE and X-ray bands in 1ES 1959+650 \citep{krawczynski2004multiwavelength}. The mechanism that causes delayed flares is unknown and a matter of debate. Two possibilities will be briefly discussed in Section \ref{sec:orphan}.

\subsection{Broadband SED modelling}
\label{sec:modeling}

\begin{table*}[htp]
\centering
\caption{Parameters for the fit to the detection-night SED shown in Fig. \ref{fig:SEDfit} with a hadronic model.  
}
\label{tab:parameters}

\begin{tabularx}{0.8\textwidth}{lXX} 
 \hline
 Parameter & Symbol & Value \\
 \hline
 Minimum electron Lorentz factor  & $\gamma_{\rm e, min}$  & $2.0 \times 10^2$  \\
  Maximum electron Lorentz factor  & $\gamma_{\rm e, max}$  & $3.8 \times 10^3$  \\
  Electron spectral index  & $q_e$  & 3.5  \\
  Escape time-scale parameter  & $\eta_{\rm esc}$  & 4.0  \\
  Magnetic field  & $B$  & 45~G \\
  Bulk Lorentz factor  & $\Gamma$  & 10  \\
  Blob radius  & $R$  & $8.5 \times 10^{15}$~cm  \\
  Accretion disk luminosity  & $L_{\rm AD}$  & $1.4 \times 10^{45}$~erg~s$^{-1}$  \\
  Minimum proton Lorentz factor  & $\gamma_{\rm p, min}$  & 1  \\
  Maximum proton Lorentz factor  & $\gamma_{\rm p, max}$  & $5 \times 10^9$   \\
  Proton spectral index  & $q_p$  & 1.9  \\
  \hline
         DERIVED QUANTITIES \\
  \hline
  Electron luminosity & $L_{\rm e}$ & $7.4  \times 10^{42}$~erg~s$^{-1}$  \\
  Proton luminosity & $L_{\rm p}$ & $8.7 \times 10^{46}$~erg~s$^{-1}$  \\
  Magnetic field luminosity & $L_{\rm B}$ & $5.5 \times 10^{46}$~erg~s$^{-1}$  \\
  Ratio of magnetic field luminosity to electron luminosity & $L_{\rm B}/L_{\rm e}$ & $7.4 \times 10^3$ \\
  Ratio of magnetic field luminosity to proton luminosity & $L_{\rm B}/L_{\rm p}$ & 0.6 \\
  Ratio of electron luminosity to proton luminosity & $L_{\rm e}/L_{\rm p}$ & $8.4 \times 10^{-5}$ \\
  \hline
\end{tabularx}
\tablefoot{All parameters provided in the first section are free while the parameters in the second section are derived quantities. $L_{\rm e}$, $L_{\rm p}$  and  $L_{\rm B}$ are all in AGN rest frame.}
\end{table*}

\begin{figure}[h]
    \centering
    \includegraphics[width=\columnwidth]{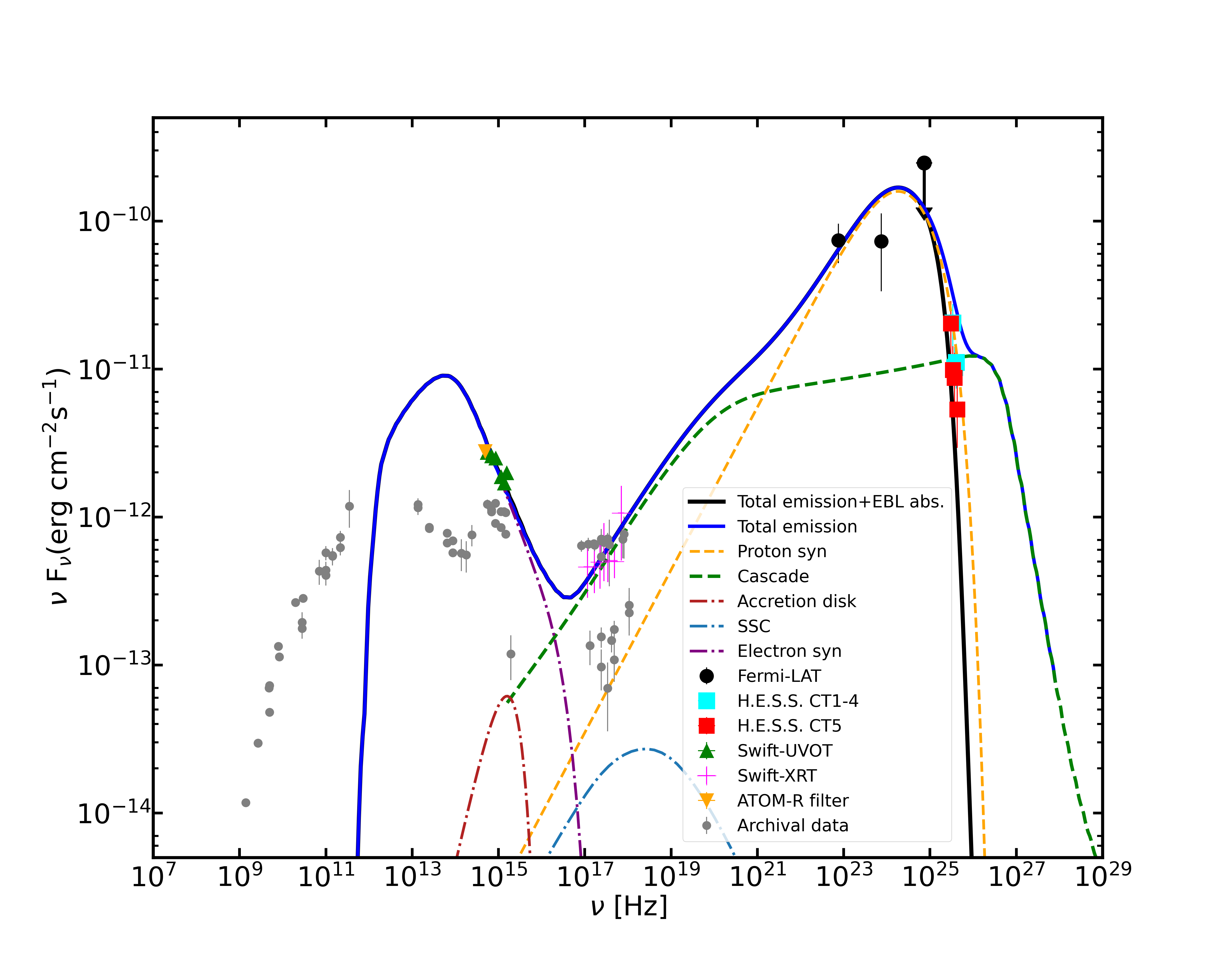}
    \caption{SED of PKS 0346-27 during the H.E.S.S. detection night of 3 November 2021, fit with a hadronic single-zone model. The individual radiation components are shown without EBL absorption, adding up to the total, intrinsic model SED shown by the solid blue curve.
    The black solid line shows the total model curve accounting for EBL absorption following the \cite{finke2010modeling} model. The {\it Fermi}-LAT emission is explained by the proton-synchrotron component while the casacade component explains the H.E.S.S. emission. The archival flux points are from \url{https://tools.ssdc.asi.it/SED/}. See Table \ref{tab:parameters} for model parameters.   }
    \label{fig:SEDfit}
\end{figure}

We compiled an SED of PKS~0346-27 using 
 observations by H.E.S.S., {\it Fermi}-LAT, ATOM and {\it Swift} XRT and UVOT, considering only the detection night. Figure \ref{fig:SEDfit} shows the  broadband SED of PKS 0346-27 modelled using a single-zone, time independent hadronic code by \cite{bottcher2013leptonic}. This model assumes a spherical emitting region of comoving radius R, filled with a tangled magnetic field B, and propagating with a bulk Lorentz factor $\Gamma$ along the jet at an angle $\theta$ with respect to the observer's line of sight, resulting in relativistic Doppler boosting by a Doppler factor $\delta = \left( \Gamma \, [1 - \beta_{\Gamma} \cos\theta] \right)^{-1}$, where $\beta_{\Gamma}$ is the normalized velocity corresponding to the Lorentz factor $\Gamma$. Primary populations of relativistic, non-thermal electrons and protons are injected into the emitting region with a power-law distribution $Q_{e/p}(\gamma_{e/p}) = Q_{0, e/p} \gamma_{e/p}^{-q_{e/p}} \; H (\gamma_{e/p}; \gamma_{\rm e/p, min},\gamma_{\rm e/p, max})$ where H is the Heaviside function defined as H = 1 if $a \le x \le b$ and H = 0 otherwise. The code reaches an equilibrium between the relativistic particle injection, radiative cooling due to various radiative processes (e.g. synchrotron radiation, photo-pion production) and escape on a timescale $t_{\rm esc}= \eta_{\rm esc} \, R/c$, where $\eta_{\rm esc}$ $\ge 1$ is a free escape-time-scale parameter. The synchrotron emission from the injected electrons forms the predominant target photon field for proton-photon interactions. \\
\indent The low energy component of the SED fit is generated by  electrons through synchrotron radiation, while proton-synchrotron emission and the synchrotron emission from secondaries produced in proton-photon (photo-pion) interactions account for the high energy component of the SED. Protons interact with the target photon field produced by the primary electrons and generate mesons ($\pi^0$, $\pi^-$ and $\pi^+$). $\pi^0$ decay into photons while charged pions decay into muons and neutrinos; the muons further decay into electrons/positrons and further neutrinos. 
\\
\indent 
Figure \ref{fig:SEDfit} shows the model SED with the EBL absorption taken into account following \cite{finke2010modeling}. Other EBL models were also used, but resulted in virtually indistinguishable VHE spectra.
The {\it Fermi}-LAT spectrum is dominated by proton-synchrotron emission, while the H.E.S.S. spectrum is dominated by photo-pion induced cascade emission. 
\\
\indent Table \ref{tab:parameters} lists all relevant parameters used to produce the SED fit.  
Note, however, that there are substantial parameter degeneracies, so the listed parameter values only represent one possible realization of a hadronic single-zone fit and should not be taken as a unique solution. Except for the redshift and the black-hole mass (see below), none of the parameters can be reliably constrained from independent observations. Therefore, they were left free with the aim of minimizing the required jet power and being in line with previous hadronic modelling results of FSRQs \citep[e.g.][]{bottcher2013leptonic}. 
\\
\indent In our hadronic SED fit, the jet power is dominated by the kinetic energy of the relativistic protons, leading to energy partition fractions of $L_B/L_e = 7.4 \times 10^{3}$ and $L_B/L_p$ = 0.6. The mass of the supermassive black hole in PKS~0346-27, $M_{\rm BH} \sim 2 \times 10^8 \, M_{\odot}$, as estimated by \cite{angioni19}, corresponds to an Eddington luminosity of $L_{\rm Edd} \sim 2.5 \times 10^{46}$~erg~s$^{-1}$. Therefore, notably, the total jet power required by our model fit exceeds the Eddington luminosity by a factor of $\sim 6$ during the H.E.S.S. detection night. However, as the modelled SED represents a short-term flare state, the Eddington limit needs to be surpassed only for the $\lesssim 1$~day duration of the flare and not persistently.  
\\
The steep optical (electron-synchrotron) spectrum necessitates a steep electron injection index of $q = 3.5$, which is significantly steeper than expected from, e.g., non-relativistic shock acceleration ($q = 2$). However, if shock acceleration is the dominant non-thermal particle-energization mechanisms, the shocks in AGN jets are likely to be at least mildly relativistic and oblique. In such a case, spectral indices significantly exceeding 2 may result from diffusive shock acceleration \citep[e.g.][]{SB12}.
\\
\indent 

We also attempted a fit with a single-zone leptonic model. However, as we will show below and in the Appendix, such a scenario would require extreme, implausible parameter choices. To illustrate the general problem with a single-zone leptonic interpretation, we start with the standard assumption that IC scattering of an external target photon field (EC = External Compton) is responsible for the $\gamma$-ray emission. In order to achieve EC emission in the Thomson regime extending into the VHE range, a low-energy target photon field is required, plausibly provided by the dusty torus of temperature $T_{\rm DT} \equiv 10^3 \, T_3$~K, with $T_3 \sim 1$ -- a few being a parameter. The $\nu F_{\nu}$ peak frequency of EC scattering of dusty torus photons (EC[DT]) by a relativistic electron population with a break / low-energy cut-off at $\gamma_b$ is given 
by $\epsilon_{\rm EC}^{\rm obs} \sim 5 \times 10^{-7} \, T_3 \, \Gamma \, \delta \, \gamma_b^2 / (1 + z)$, if Compton scattering occurs in the Thomson regime, which we will verify a posteriori. Given the required steep electron spectrum, this peak frequency must be located above the {\it Fermi}-LAT energy range, in order not to over-produce the GeV $\gamma$-ray flux. Assuming the EC(DT) peak frequency to be at a dimensionless energy $\epsilon_{\rm EC}^{\rm obs} \equiv E_{\rm EC}^{\rm obs}/(m_e c^2) \sim 10^5$, we require a break electron energy of $\gamma_b \sim 6 \times 10^4 \, (\Gamma_1 \, \delta_1 \, T_3)^{-1/2}$, where we use the standard nomenclature of $Q_i = Q/(10^i [\rm c.g.s])$. As the target photon energy in the co-moving frame is $\epsilon'_t \sim 5 \times 10^{-6} \, \Gamma_1 \, T_3$, we find that $\epsilon'_t \gamma_b \sim 0.3 \, T_3^{1/2} \, (\Gamma_1 / \delta_1)^{1/2}$, which confirms our assumption that Compton scattering occurs in the Thomson regime at the peak of the $\gamma$-ray spectrum, as long as $\delta \approx \Gamma$. Now, the steep optical -- UV spectrum indicates that the synchrotron peak is located at $\nu_{\rm sy}^{\rm pk} \lesssim 5 \times 10^{14}$~Hz, leading to a magnetic-field estimate of $B \lesssim 6 \, \Gamma_1 \, T_3$~mG, much lower than the $\sim $~G magnetic fields typically found in SED modeling of FSRQs \citep[e.g.][]{aleksic2011magic, zacharias2017cloud, angioni19}. The Poynting-flux power carried by the jet of cross-sectional radius $R_B \equiv 10^{16} \, R_{16}$~cm, is $L_B \lesssim 1.4 \times 10^{39} \, \Gamma_1^4 \, T_3^2 \, R_{16}^2$~erg~s$^{-1}$. As the power in relativistic electrons in the jet has to be at least as large as the radiated power (dominated by the $\gamma$-ray emission with $(\nu F_{\nu})^{\rm EC} \sim 10^{-10}$~erg~cm$^{-2}$~s$^{-1}$), we have a lower limit on $L_e \gtrsim 4 \pi \, d_L^2 \, (\nu F_{\nu})^{\rm EC} / \Gamma^2 \sim 5 \times 10^{46} \, \Gamma_1^{-2}$~erg~s$^{-1}$, where we used a luminosity distance of $d_L \sim 2 \times 10^{28}$~cm. Thus, the jet would need to be extremely far out of equipartition, dominated by electron kinetic energy, by a factor $L_B/L_e \lesssim 3 \times 10^{-8} \, \Gamma_1^6 \, R_{16}^2 \, T_3^2$, while the $\lesssim$~day-scale variability observed ($t_{\rm var} \equiv 1 \, t_d$~day) constrains the size of the emission region to $R_{16} \lesssim 1.3 \, t_d \, \delta_1$. Thus, for any plausible choice of parameters, the jet would be unreasonably strongly kinetic-energy dominated. 

Another problem becomes obvious when evaluating the co-moving synchrotron photon energy density, $u'_{\rm sy} \sim d_L^2 \, (\nu F_{\nu})^{\rm sy} / (R_B^2 \, c \, \delta^4) \sim 0.13 \, R_{16}^{-2} \, \delta_1^{-4}$~erg~cm$^{-3}$. Comparing this to the magnetic-field energy density, $u'_B \lesssim 10^{-7} \, \Gamma_1^2 \, T_3^2$~erg~cm$^{-3}$, we find that a ratio of SSC to synchrotron peak fluxes, $(\nu F_{\nu})^{\rm SSC} / (\nu F_{\nu})^{\rm sy} \approx u'_{\rm sy} / u'_B \sim 3 \times 10^6 \, R_{16}^{-2} \, \delta_1^{-4} \, \Gamma_1^{-2} \, T_3^{-2}$ would result. Given that this SSC dominance should not exceed the observed Compton dominance of $\sim 10$, we can infer that $\Gamma \sim \delta \gtrsim 80 \, (R_{16} \, T_3)^{-1/3}$, which is unusually large compared to the typically inferred Doppler and bulk Lorentz factors $\sim$ 10 -- 50 \citep{lister2016agn},  although \cite{2021ApJ...923...67H} reported Doppler factors of $>$ 100 for several blazars studied in the radio band.  The above estimates illustrate that a single-zone leptonic model has great difficulties producing the observed flare-state SED with plausible parameter values.

An example of a fit attempt using parameters similar to those constrained above, is presented in the Appendix, illustrating that a single-zone leptonic model with $\Gamma = \delta =80$ is able to reproduce the flare-state SED of PKS~0346-27, however, requiring extreme parameters: exceedingly large Doppler and bulk Lorentz factors, an  unusually small magnetic field of 48 mG (compared to SED fit results for other FSRQ-type blazars), and particle energy dominating over magnetic-field energy by over 5 orders of magnitude. We therefore disfavored such a scenario. 
Thus, when restricting the model consideration to a basic one-zone model, the hadronic model discussed above was considered the most plausible interpretation. Leptonic multi-zone models with two or more zones contributing jointly to produce a broad high-energy emission component can, of course, not be ruled out, but their consideration is beyond the scope of this paper.
As the hadronic single-zone model is the model with the fewest additional parameters beyond a single-zone leptonic model and it provides a plausible SED fit, we consider it preferred.

Difficulties with SED fitting using a single-zone leptonic model have also been found in other IACT-detected FSRQs. For example, the steep optical -- UV spectrum of 3C279, compared to the hard and broad $\gamma$-ray SED, extending into the VHE regime, made such a single-zone leptonic fit implausible, leading to a strong preference for a hadronic interpretation \citep{boettcher2009implications}. In the case of PKS 1510-089 \citep{2023ApJ...952L..38A}, the uncorrelated variability in 2021-22, with the X-ray and VHE emissions remaining at almost unchanged levels, while the optical and {\it Fermi}-LAT $\gamma$-ray emissions dropped abruptly in 2021, provided a strong preference for two-zone interpretation.

We note, however, that the SEDs of other VHE-detected FSRQs could be plausibly reproduced with single-zone leptonic models: The SEDs of PKS~1441+25 \citep{Ahnen15_PKS1441,Abeysekara15_PKS1441}, PKS  1222+216, and TON 599 \citep{Adams22} could all be modelled with single-zone leptonic scenarios, and even the SED of PKS 1510-089 during an outburst in 2015, with less unusual variability features than the 2021-2022 period mentioned above, could be well represented with such a model \citep{Ahnen17_PKS1510}. In most of these cases, there is evidence for synchrotron emission making a significant contribution to the X-ray emission as well as significantly harder Fermi-LAT (and in  most cases also harder IR -- optical -- UV synchrotron) spectra than found in PKS 0346-27, which indicates harder electron spectra, extending to higher electron energies, facilitating a leptonic high-energy-emission interpretation.

\subsection{Possible causes of TeV -- GeV flare delays}
\label{sec:orphan}

The possible delay of the VHE flare on MJD 59521.9, $\sim 2$~days after the peak of the {\it Fermi}-LAT flare, deserves special consideration. While no contemporaneous H.E.S.S. observations had been possible during the HE flare and therefore we can not exclude a simultaneous HE + VHE $\gamma$-ray flare, the VHE flare during the H.E.S.S. detection night clearly appears as a VHE flare without simultaneous counterpart in HE $\gamma$-rays. Various models have been suggested for delayed $\gamma$-ray flares, including the class of synchrotron mirror models, in which synchrotron radiation from the dominant jet emission region is reflected off stationary (or non-relativistically moving) clouds / sheaths near the jet trajectory and re-enters the jet to act as an enhanced target photon field for Compton scattering in leptonic emission scenarios \citep{GM96,BD98,Bednarek98,Vittorini14,Tavani15,McDonald15,McDonald17,Boettcher21} or for photo-pion production in hadronic scenarios \citep{bottcher2005hadronic,2018heas.confE..24O}. The latter model may lead to an orphan VHE flare due to enhanced $\gamma$-ray production from pion decay and subsequent electromagnetic cascades, while the HE $\gamma$-ray emission is dominated by proton-synchrotron emission (as in the case of our SED fit presented above), not significantly affected by the enhanced external radiation field. Such a scenario would require the reflecting material to be located at a distance of $R_m \sim 2 \Gamma^2 \, c \Delta t / (1 + z)$, where $\Delta t \sim 2$~days is the delay between the primary HE $\gamma$-ray flare and delayed VHE flare. With the value of $\Gamma = 10$ adopted for our hadronic SED fit, this would yield a distance of $R_m \sim 0.2$~pc, a plausible location of an isolated broad-line-region (BLR) cloud. Detailed modelling of this flare with such a hadronic synchrotron mirror model is beyond the scope of this  paper and is therefore left to future work. 

An alternative explanation could be a finite acceleration time of protons producing the VHE $\gamma$-ray emission via proton synchrotron radiation \citep[see, e.g.][for an application of a similar scenario to the delayed VHE emission from the recurrent nova RS Ophiuchi]{HESS22RSOph}. The proton acceleration time scale for a generic acceleration process for the highest-energy protons of energy $\gamma_{\rm p,max} = 5 \times 10^{9} \, (B / 45 {\rm G})^{-1/2} \, \delta_1^{-1/2}$ may be parameterized as 

\begin{equation}
t'_{\rm acc} \equiv \eta_a {\gamma_{\rm p, max} \, m_p \, c \over e \, B}  \approx 1.2 \times 10^4 \, \eta_a \, \left( {B \over 45 \, {\rm G}} \right)^{-3/2} \, \delta_1^{-1/2} \, {\rm s}
\label{tacc}
\end{equation}
with an acceleration efficiency parameter $\eta_a \ge 1$, where the prime indicates the time in the co-moving frame of the emission region. For this acceleration time to equal the co-moving-frame time delay between the {\it Fermi}-LAT and H.E.S.S. flares, $\Delta t' = \Delta t^{\rm obs} \, \delta / (1 + z) \approx 9 \times 10^5 \, \delta_1$~s, we require an acceleration efficiency factor of 

\begin{equation}
\eta_a \approx 75 \, \left( {B \over 45 \, {\rm G}} \right)^{3/2} \, \delta_1^{3/2},
\label{eta_a}
\end{equation}
a plausible value for a moderately efficient acceleration process. The synchrotron cooling time scale for protons of energy $\gamma_{\rm p, max}$ is

\begin{equation}
t'_{\rm sy} = \left( {m_p \over m_e} \right)^3 {6 \, \pi \, m_e \, c^2 \over c \, \sigma_T \, B^2 \, \gamma_{\rm p,max}} \approx 4.7 \times 10^5 \, \left( {B \over 45 \, {\rm G}} \right)^{-3/2} \,\delta_1^{1/2} \, {\rm s}
\label{tsy}
\end{equation}
which is of the same order of magnitude as the acceleration time scale from Eq. \ref{tacc} (with $\eta_a = 75$), so that radiative cooling shuts off the acceleration when protons reach the energy $\gamma_{\rm p,max}$. The  Larmor radius of protons of such energy is $r_L \approx 3.5 \times 10^{14} \, (B / 45 \, {\rm G})^{-3/2} \, \delta_1^{-1/2}$~cm, smaller than the size of the emission region used for our SED modelling. Hence, the protons required  for producing the VHE $\gamma$-ray proton-synchrotron emission can well be confined within the emission region by a magnetic field of $\sim 45$~G. 

This scenario, however, faces two problems: First, it does not explain why the {\it Fermi}-LAT flux has decreased back to quiescent levels at the time of the H.E.S.S. flare, as the acceleration process is still active and the synchrotron cooling time scale for protons producing the HE $\gamma$-ray emission is longer than 2~days in the observer's frame. Second, even the cooling time scale for protons producing the VHE emission is $\approx 2$~days, so that additional factors --- possibly a swing of the motion of the emission region away from the line of sight \citep[e.g.][]{Britzen17,Britzen23} or a decreasing magnetic field on a time scale of $\sim 1$~day in the observer's frame \citep[see, e.g.,][for a study of magnetic-field changes on the long-term variability of blazars]{Thiersen22,Thiersen24} --- must be invoked in order to explain the $< 1$~day duration of the VHE flare.

\section{Summary and conclusions}

In this paper, we report the VHE $\gamma$-ray detection of the LSP  blazar PKS~0346-27 by H.E.S.S. on 03 November 2021, as well as contemporaneous multi-wavelength observations. At the time of the announcement of this detection \citep{Wagner21}, this was the most distant  VHE-detected blazar at $z = 0.991$ \citep[this has now been superseded by OP 313 at $z = 0.997$, detected by the CTAO LST-1, ][]{LSTOP313}. Day-scale $\gamma$-ray variability was found. H.E.S.S. detected the source in only one night, $\sim 2$~days after the peak of a prominent HE $\gamma$-ray flare detected by {\it Fermi}-LAT, which had subsided at the time of the H.E.S.S. detection. The broadband SED on the day of the H.E.S.S. detection could be satisfactorily modelled with a proton-synchrotron dominated single-zone hadronic model with temporarily super-Eddington jet power. In the framework of such a hadronic model, the potentially delayed VHE flare could possibly be explained with a hadronic synchrotron mirror model or a scenario of gradual proton acceleration, although the latter scenario faces significant challenges. Detailed model simulations of such scenarios are left to future work. 
 Alternatively, a fit with a single-zone leptonic model is, in principle, also possible, but it requires extreme Doppler and bulk Lorentz factors of $\sim 80$, an unusually low magnetic field (compared to SED fitting results of other FSRQs), and an energy ratio between relativistic electrons and magnetic fields far out of equipartition by more than 5 orders of magnitude. We therefore favor the hadronic model interpretation. Multi-zone models are another alternative that might provide a satisfactory SED fit, but their exploration is beyond the scope of this paper. 
\\
{\tiny$\it Acknowledgements$. The support of the Namibian authorities and of the University of Namibia in facilitating the construction and operation of H.E.S.S. is gratefully acknowledged, as is the support by the German Ministry for Education and
Research (BMBF), the Max Planck Society, the German Research Foundation
(DFG), the Helmholtz Association, the Alexander von Humboldt Foundation,
the French Ministry of Higher Education, Research and Innovation, the Centre National de la Recherche Scientifique (CNRS/IN2P3 and CNRS/INSU), the
Commissariat à l’Énergie atomique et aux Énergies alternatives (CEA), the U.K.
Science and Technology Facilities Council (STFC), the Irish Research Council (IRC) and the Science Foundation Ireland (SFI), the Knut and Alice Wallenberg Foundation, the Polish Ministry of Education and Science, agreement
no. 2021/WK/06, the South African Department of Science and Technology and
National Research Foundation, the University of Namibia, the National Commission on Research, Science and Technology of Namibia (NCRST), the Austrian
Federal Ministry of Education, Science and Research and the Austrian Science
Fund (FWF), the Australian Research Council (ARC), the Japan Society for the
Promotion of Science, the University of Amsterdam and the Science Committee of Armenia grant 21AG-1C085. We appreciate the excellent work of the
technical support staff in Berlin, Zeuthen, Heidelberg, Palaiseau, Paris, Saclay,
Tübingen and in Namibia in the construction and operation of the equipment.
This work benefited from services provided by the H.E.S.S. Virtual Organisation, supported by the national resource providers of the EGI Federation.}

\bibliography{0346}

\begin{appendix}

\section{Leptonic model fitting attempts}

As mentioned in the main text, we attempted to fit the SED of the detection-night observations with the leptonic model described in \cite{bottcher2013leptonic}. The setup is similar to the one described in Section \ref{sec:modeling}, but not including radiative signatures of ultra-relativistic protons. In addition, high-energy emission is produced by Compton upscattering of synchrotron emission (SSC) and external radiation fields (EC). The external radiation fields include the direct accretion-disk emission (EC[disk]) and a thermal blackbody radiation field, representative of the dusty torus (EC[DT]), modelled as isotropic in the AGN rest frame. 
 
\begin{figure}[h]
    \centering
    \includegraphics[width=\columnwidth]{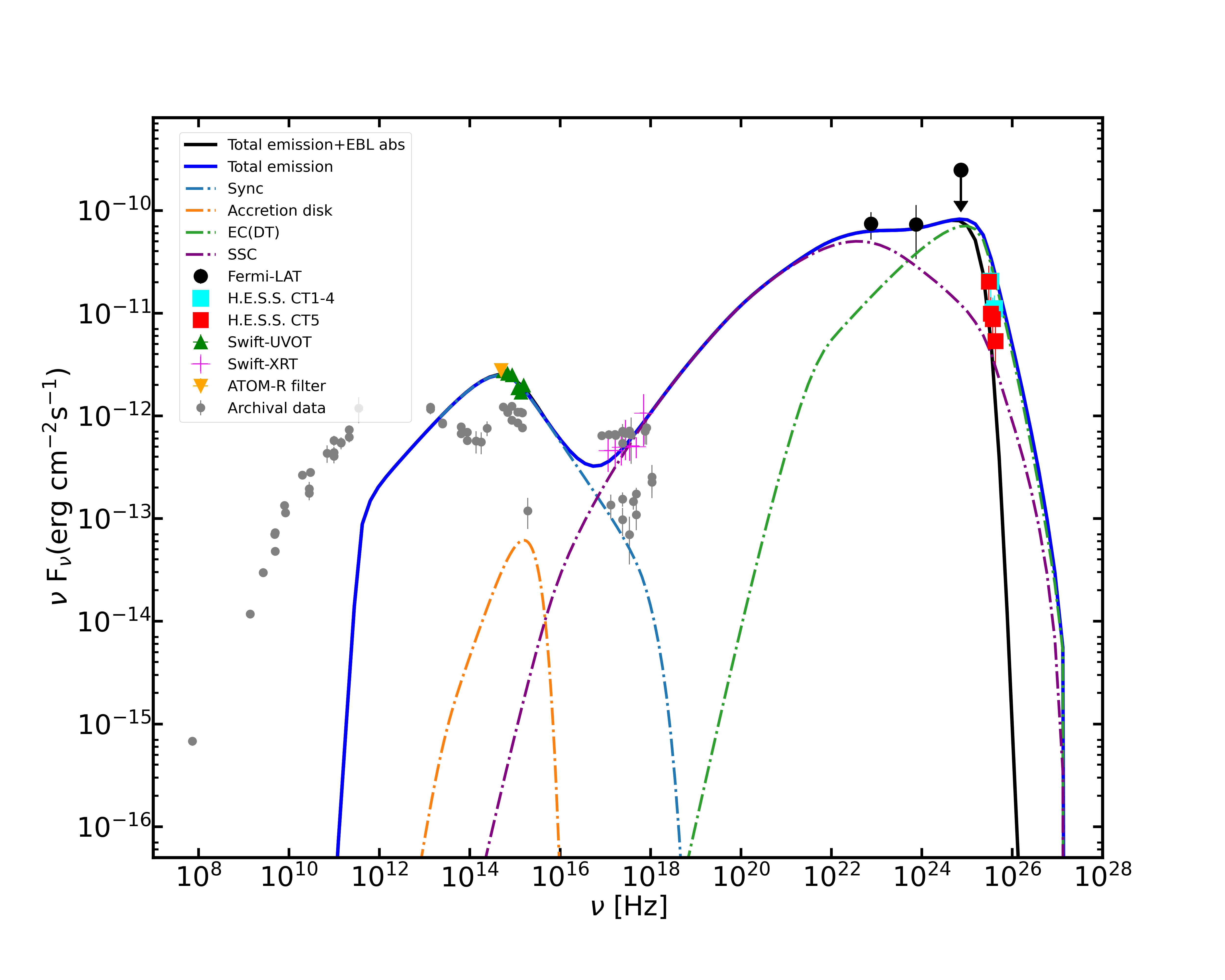}
    \caption{Representative attempt of single-zone leptonic model fit to the SED of PKS 0346-27 during the H.E.S.S. detection night of 3 November 2021. The individual radiation components are shown without EBL absorption, adding up to the total, intrinsic model SED shown by the solid blue curve. The black solid line shows the total model curve accounting for EBL absorption following the \cite{finke2010modeling} model. The archival flux points are from \url{https://tools.ssdc.asi.it/SED/}. See Table \ref{tab:leptonic_parameters} for model parameters. }
    \label{fig:leptonicSEDfit}
\end{figure}

\begin{figure}[h]
    \centering
    \includegraphics[width=\columnwidth]{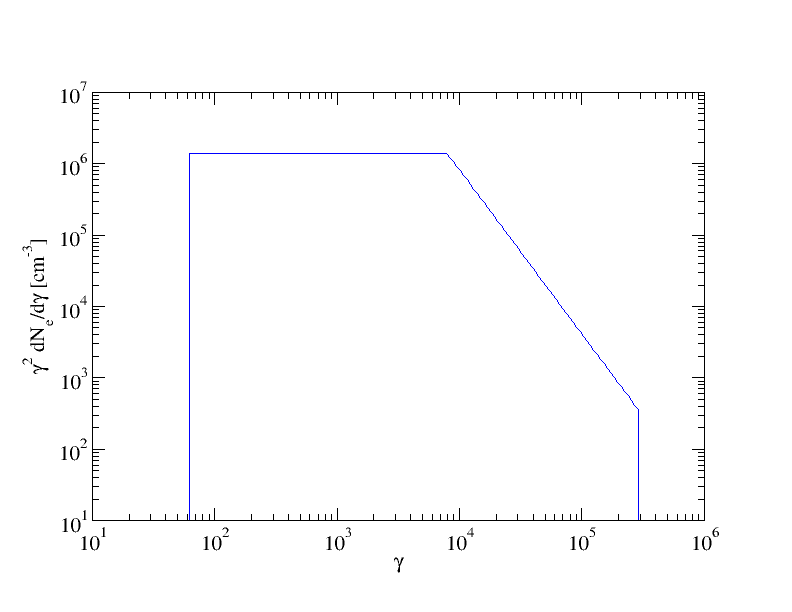}
    \caption{Radiating electron spectrum from the leptonic SED fit shown in Fig. \ref{fig:leptonicSEDfit}. }
    \label{fig:espectrum}
\end{figure}

Fig. \ref{fig:leptonicSEDfit} shows a representative fitting attempt, utilizing parameter values similar to those estimated in the main text (with $\Gamma = \delta = 80$) and listed in Table \ref{tab:leptonic_parameters}.
The radiating (equilibrium) electron spectrum from that model fit is shown in Fig. \ref{fig:espectrum}.

\begin{table*}[ht]
\caption{Parameters for the representative leptonic-model fit attempt to the detection-night SED shown in Fig. \ref{fig:leptonicSEDfit}. }
\label{tab:leptonic_parameters}
\begin{tabularx}{0.8\textwidth}{lXX} 
 \hline
 Parameter & Symbol &   Value \\
 \hline
 Minimum electron Lorentz factor  & $\gamma_{\rm e, min}$  & $8.0 \times 10^3$  \\
  Maximum electron Lorentz factor  & $\gamma_{\rm e, max}$  & $3.0 \times 10^5$  \\
  Electron spectral index  & $q_e$  & 3.3  \\
  Escape time-scale parameter  & $\eta_{\rm esc}$  & $1.0 \times 10^3$  \\
  Magnetic field  & $B$  & 0.048~G \\
  Bulk Lorentz factor  & $\Gamma$  & 80  \\
  Blob radius  & $R$  & $2.4 \times 10^{15}$~cm  \\
  Accretion disk luminosity  & $L_{\rm AD}$  & $1.4 \times 10^{45}$~erg~s$^{-1}$  \\
  Distance from black hole & $z_0$ & 1.0 pc \\
  Ext. rad. field BB temperature & $T_{BB}$ & $10^3$~K \\
  Ext. rad. field energy density & $u_{\rm ext}$ & $4.0 \times 10^{-7}$~erg~cm$^{-3}$ \\
  \hline
  \\
         DERIVED QUANTITIES \\
  \hline
  \\
  Electron luminosity & $L_{\rm e}$ & $4.1 \times 10^{46}$~erg~s$^{-1}$  \\
  Magnetic field luminosity & $L_{\rm B}$ & 
  $3.2 \times 10^{41}$~erg~s$^{-1}$  \\
  Ratio of magnetic field luminosity to electron luminosity & $L_{\rm B}/L_{\rm e}$ & $7.7 \times 10^{-6}$ \\  
\hline
\end{tabularx}
\end{table*}

While, in principle, a satisfactory SED fit is possible with this setup, the required choices of parameters is extreme: 
 In particular, the system is more than 5 orders of magnitude out of equipartition (particle dominated), which poses problems of particle confinement and the source of energy powering particle acceleration. Also, the magnetic field of 48 mG is much lower than what is usually found in SED modeling of FSRQ-type blazars and the large Doppler and bulk Lorentz factor of 80 appear uncomfortably high, although such values have been inferred from radio observations in a few cases.
We therefore conclude that a single-zone leptonic model is dis-favored compared to the hadronic model fit presented in the main text, which could be achieved with much more natural parameter choices.

\end{appendix}

\end{document}